\documentclass[prl,twocolumn,superscriptaddress,showpacs]{revtex4}
\usepackage{bm}
\begin{document}
\def\btt#1{{\tt$\backslash$#1}}
\def\tr{{\rm\,tr\,}}
\newcommand{\bra}[1]{\langle #1|}
\newcommand{\ket}[1]{|#1\rangle}
\newcommand{\braket}[2]{\langle #1|#2\rangle}
\newcommand{\half}{\textstyle{\frac{1}{2}}}
\newcommand{\beq}{\begin{equation}}
\newcommand{\eeq}{\end{equation}}
\newcommand{\bdis}{\begin{displaymath}}
\newcommand{\edis}{\end{displaymath}}
\newcommand{\bea}{\begin{eqnarray}}
\newcommand{\eea}{\end{eqnarray}}
\newcommand{\barr}{\begin{array}}
\newcommand{\earr}{\end{array}}
\newcommand{\beas}{\begin{eqnarray*}}
\newcommand{\eeas}{\end{eqnarray*}}
\title{Nonlinear elastodynamics and energy transport in biopolymers}
\author{Radha Balakrishnan}
\affiliation{The Institute of Mathematical Sciences, Chennai  600 113,
India}
\email {radha@imsc.res.in}
\author{Rossen Dandoloff}
\affiliation{Laboratoire de Physique Th\'eorique et Mod\'elisation,
Universit\'e de Cergy-Pontoise, F-95302 Cergy-Pontoise, France}
\email{rossen.dandoloff@ptm.u-cergy.fr}
\begin{abstract}
We present an exact analysis of the intrinsic dynamics of two types of
biopolymers, described by the wormlike chain (WLC) and the
wormlike rod chain (WLRC) models, respectively.  We map each model
to a classical field theory for an appropriate unit vector field
defined on the space curve formed by the axis of the biopolymer.
Conformation dynamics and
energy transport are shown to be described, respectively, by the
soliton-bearing Landau-Lifshitz equation and the nonlinear Schr\"odinger
equation. This enables us to identify
the soliton representing
the ``conformon'' hypothesized in  biopolymers.
\end{abstract}

\pacs{02.40Hw, 05.45Yv, 87.10.+e, 87.15He}

\maketitle

In the years since the  discovery of biological
polymers such as proteins and
DNA, it has been recognized\cite{vino} that the geometric
shapes and topological properties of these
long chain molecules play a significant role
during processes such as replication and transcription.
Experimental techniques capable of studying the conformation and
elastic properties of a {\it single} polymer by bending or twisting
it are also available\cite{smit}. Semiflexible biopolymers such as
actin have only a bending energy, and are described by the well-known
wormlike chain (WLC) model\cite{goldstein} with a single elastic constant,
the bending modulus. In contrast, DNA with its double-helix
structure is described by the wormlike rod chain (WLRC) model
\cite{fain} with an additional elastic constant, the twist rigidity.
Although considerable work has been done on various equilibrium
properties of both these elastic models, their
intrinsic dynamical properties have
not been studied so far. The latter play a crucial role in
the mechanisms of energy and information propagation along a
biopolymer, an issue of vital interest to biologists,
chemists and physicists alike.

In this Letter we present, for the first time, a formalism to study
the exact intrinsic dynamics of a polymer in the WLC and WLRC models.
Several new results are obtained.
By modeling the polymer as a slender
elastic rod whose central axis is a space curve,
the appropriate moving frame comprising the unit tangent,
normal and binormal vectors is
constructed for each model. We show that the WLC energy
is a certain classical field theory Hamiltonian
for the {\it tangent} vector, while the WLRC energy is an analogous
Hamiltonian for the {\it normal} vector field.
Using a variational principle  for the Hamiltonian, and imposing
the unit vector constraint, the equilibrium conformations
of the polymer are first found.
Next, by setting the time derivative of
the Hamiltonian to zero, the time evolutions of the vector
fields consistent with their corresponding equilibrium
conformations are found for
the two models. It turns out that the equation of motion of the
relevant vector field in each model
resembles that of the precession of a magnetic moment\cite{gold},
i.e., a ``spin'', in a self-induced ``magnetic field''.
We also show  that the dynamical equation in
each of the models can be derived as Hamilton's equations
from the respective elastic Hamiltonians,
by using the $z$-component of the vector field
and its azimuthal angle
as the canonically conjugate variables.
As is well known, these variables
give rise to the correct Poisson brackets  for the ``spin''
components. Remarkably enough,
the time evolution equations for the tangent vector
in the WLC model and the normal vector in the WLRC model
have the same form as the well-known  Landau-Lifshitz equation.
The latter is completely integrable by
the inverse scattering transform method.
We show that the  {\it writhe} and the {\it twist}
of the vector field in each model are conserved
quantities under time evolution.
By White's theorem\cite{whit}, this leads to the
conservation of its {\it linking number} as well.

The non-extensibility
of the polymer during time evolution entails
compatibility conditions
on the frame fields of the
polymer. Imposing these leads
to the completely
integrable nonlinear Schr\"odinger equation (NLS) \cite{zakh}
for the energy density in each model. This shows that
energy can propagate along the polymer like a localized
wave packet, in both types of biopolymers.

Using the one-soliton solution of the NLS equation,
we are able to identify the curvature and torsion
functions in the two models. From these expressions
we find that, in the WLC model,
the soliton corresponds to a
localized ``loop'' propagating along the
semiflexible polymer, while in
the WLRC model (applicable to DNA), it is
possible to have a localized ``supercoiled''
entity propagating along the polymer,
for appropriate values of the width and
velocity parameters of the soliton.
We interpret our solutions
as the explicit manifestation  of the
so-called ``{\it conformon}'', an entity
that has been hypothesized  by various
authors \cite{volk,gree,keme}
to play an important role  in
biology -- for example, these are localized packets
of conformational strain on mitochondria.

In the  context of an
$\alpha$-helical protein, the  NLS soliton first found by
Davydov\cite{davy} using a {\it quantum mechanical} lattice model is
well studied.
However, it has been noted\cite{scot}
that the  Davydov soliton is
essentially a {\it classical} object. Our analysis leads directly to
a classical soliton.
It also enables us to suggest a mechanism for charge transport along the polymer,
in the light of the results of Goldstone and Jaffe\cite{goldstone}
on the curvature-induced localization of a quantum particle propagating
in a thin curved tube.\\

\noindent
{\it The WLC model}:\,
A semiflexible biopolymer is represented  as a slender elastic
rod with a single elastic modulus characterizing
the bending energy of the polymer \cite{goldstein}. In general, the axis of
the polymer is a space curve.
The unit tangent $\mathbf{t}$, normal $\mathbf{n}$ and binormal
$\mathbf{b}=\mathbf{t}\times\mathbf{n}$ comprise a frame on the space
curve. They satisfy the well known Frenet-Serret
equations\cite{vais}
$\mathbf{t}_s=\kappa\mathbf{n}\,,\,
\mathbf{n}_s = -\kappa\mathbf{t}+\tau{\mathbf
b}\,,\,\mathbf{b}_s
=-\tau\mathbf{n}\,.$
Here $s$ is the arclength and the subscript denotes the
corresponding derivative.
As usual, ${\kappa}$ is the curvature
and ${\tau}$ the torsion of the curve.
The Frenet-Serret equations can be written in the compact
form $\mathbf{F}_s=\bm{\Omega}\times\mathbf{F}$ where
$\mathbf{F}$ stands for $\mathbf{t}$, $\mathbf{n}$ or
$\mathbf{b}$, and $\bm{\Omega}=
\tau\,\mathbf{t}+\kappa\,\mathbf{b}$.
Thus, as one moves along the curve, the frame rotates with a
``spatial angular velocity'' $\bm{\Omega}$.

Consider a polymer with a constant disc-shaped cross-section.
We may imagine the polymer to be formed by the tumbling of this
disc through space \cite{goldstone}
without twisting about its axis,
thereby implying that its angular velocity should always lie
in the plane of the disc. This amounts to
the Fermi-Walker parallel transport of the triad,
as has been recognized in another context\cite{dando}.
In order to satisfy this condition, we introduce new unit vectors
$ \mathbf{N}=\mathbf{n}\,\cos\,\alpha (s)
+\mathbf{b}\,\sin\, \alpha (s)$ and $
\mathbf{B}=-\mathbf{n}\,\sin\, \alpha (s)
+ \mathbf{b}\,\cos\, \alpha (s)\,$,
where $\alpha (s)=-\int_{-\infty}^{s}\tau(s')\,ds'.$ Then
the triad comprising  $(\mathbf{t},\mathbf{N},\mathbf{B})$
can be shown to satisfy the Fermi-Walker (FW)  equations
$\mathbf{t}_s=\kappa\,
(\mathbf{N}\,\cos\,\alpha-\mathbf{B}\,\sin\,\alpha)\,,\,
\mathbf{N}_s=-\mathbf{t}\,\kappa\,\cos\,\alpha\,$, and $\mathbf{B}_s
=\mathbf{t}\,\kappa\,\sin\,\alpha\,$.
These equations lead to
an angular velocity  $\bm{\omega}=
\kappa(\mathbf{N}\,\sin\, \alpha
+\mathbf{B}\, \cos\,\alpha)$  for the triad.
The corresponding Hamiltonian is therefore of the form
\beq
H=\frac{A}{2}\int {\mathbf \omega}^2ds=
\frac{A}{2}\int \kappa^2 \,ds=\frac{A}{2}\int \mathbf{t}_s^2\,ds,
\label{wlce}
\eeq
where $A$ is a constant. This is precisely the expression written
down for the elastic energy of the rod in the WLC model,
with $A$ standing for the bending modulus.

We first obtain the minimum energy conformations of the WLC.
We scale out $A$ and use the variational principle $\delta H=0$ on
the Hamiltonian
\beq
H=\frac{1}{2}\int \{ {\mathbf{t}_s}^2-\beta(\mathbf{t}^2-1) \}\,ds
\label{var}
\eeq
where $\beta$ is the Lagrange multiplier
corresponding to the unit vector field
constraint $\mathbf{t}^2=1$. A short calculation yields
\beq
\mathbf{t}\times\mathbf{t}_{ss}=0\,,
\label{equilt}
\eeq
whose solutions represent the static conformations sought.
To find the dynamics of the polymer, we set $\mathbf{t} =
\mathbf{t}(s,u)$ in Eq. (\ref{wlce}) where $u$ denotes the time.
The energy conservation condition $dH/du = 0$ then leads to
$\int\mathbf{t}_{su}\cdot\mathbf{t}_s\,ds=0\,$, which yields,
on integrating by parts,
$\int\mathbf{t}_{ss}\cdot\mathbf{t}_{u}\,ds=0$.
In turn, this gives $\mathbf{t}_u=\mathbf{V}\times \mathbf{t}_{ss}$, where
$\mathbf{V}$ is an arbitrary vector. The static solution of this
equation is $\mathbf{V}\times\mathbf{t}_{ss}=0$.
A comparison of this with Eq.(\ref{equilt})
immediately suggests the simplest possible choice, namely,
$\mathbf{V}=\mathbf{t}$. The corresponding time
evolution for $\mathbf{t}$ is then given by
\beq
\mathbf{t}_u=\mathbf{t}\times\mathbf{t}_{ss}\,.
\label{LL}
\eeq
This equation resembles that\cite{gold} for
the precession of a magnetic moment or spin
in a self-induced magnetic field. To make this connection more
precise, we must ask whether Eq. (\ref{LL}) can be obtained
directly from the WLC elastic Hamiltonian in Eq. (\ref{wlce}).
Now, the  unit vector $\mathbf{t}$ may be
parametrized as $\mathbf{t}=
(\sin\,\theta \,\cos\,\phi,\sin\,\theta \,\sin\,\phi,
\cos\,\theta)$. If $\mathbf{t}$ is to be a classical
spin vector, its cartesian components must satisfy
the standard Poisson bracket relations
$\{ t^{\alpha}, t^{\beta} \} =
\epsilon^{\alpha\beta\gamma} t^{\gamma}$. (In turn, this is
ensured if we have
$\{\phi, \cos\,\theta \} =1$.) With this input for the canonical
Poisson brackets, we can show that Eq. (\ref{LL}) can be
obtained from the equation of motion
$t_{u}^{\alpha}= \{ t^{\alpha}, H \}$ corresponding
to the WLC elastic Hamiltonian.
Further, the canonical Poisson bracket relations also enable us to
show that the writhe of $\mathbf{t}$, defined by
$Wr=(2\pi)^{-1}\int \phi_s\,(1-\cos\,\theta)\,ds\,$,
is a conserved quantity,
i.e., $\{ H, Wr \}=0$. The writhe is a measure of the
chirality of the curve.

We note that Eq. (\ref{LL}), which is called the
Landau-Lifshitz (LL) equation in the spin context,
is satisfied by the {\it tangent} $\mathbf{t}$
along the axis of the polymer in the WLC model.
The LL equation appears as a spin evolution
equation in the dynamics
of a continuum Heisenberg ferromagnetic chain. It
is a completely integrable equation
with strict soliton solutions\cite{takh}
and an infinite number of constants of motion.
Next,  we use  the FW equations for
$(\mathbf{t},\mathbf{N},\mathbf{B})$
 in Eq. (\ref{LL}), and impose the compatibility conditions
$\mathbf{t}_{su}=\mathbf{t}_{us}\,, \,\mathbf{B}_{su}=
\mathbf{B}_{us}\,$. After some algebra, we get
$ \kappa_u=-(\kappa\,\tau)_s-\kappa_s\,\tau$ and $\tau_u=
(\kappa^{2}/2+ \kappa_{ss}/\kappa-\tau^2)_s\,$.
These equations are just the imaginary and real parts\cite{laks} of
the nonlinear Schr\"odinger equation (NLS)
\beq
i\psi_u+\psi_{ss}+\frac{1}{2}|\psi|^2\,\psi=0,
\label{nls}
\eeq
 when $\psi=\kappa\,
\exp\,(i\int \tau\, ds)$ is  the
well-known Hasimoto function\cite{hasi}.
(In the present context, the compatibility conditions represent
the physical property of non-extensibility of the polymer.)
Thus $|\psi|^2 = \kappa^2\,$, the energy density of the polymer.
From the equation for $\tau_u\,$,
we can easily show that
the polymer evolves in such a way as to conserve its
total twist, namely, $Tw=\int \tau \,ds$.
Since we have already shown that the writhe  $Wr$ is  conserved,
it follows from White's theorem \cite{whit} $Tw+Wr=Lk$,
that the linking number $Lk$
of the WLC polymer is also conserved.

The one-soliton solution of Eq. (\ref{nls}) is given by
\beq
\psi (s,u) =p_1\,{\rm sech}\,\xi_1 \,\exp\,(i\xi_2)\,,
\label{nlss}
\eeq
with $\xi_1=p_1 (s-v_1u)$ and $\xi_2=
\frac{1}{2}v_{1}s -\frac{1}{4}(v_1^2-p_{1}^{2})u$,
where the amplitude $p_{1}$ and the
velocity $v_1$ of the envelope
of the solution are independent parameters.
The  energy density is
$|\psi|^2=p_{1}^{2}\,{\rm sech}^{2}
\,\xi_1$. This represents an energy lump propagating
with velocity $v_1$ along the polymer. For this
solution the curvature
$\kappa=p_1\,{\rm sech}\,\xi_1$
and the torsion $\tau=\frac{1}{2}v_{1}$.
It is clear from these expressions that this
solitonic excitation represents
a localized deformation of the curvature, with localization
length $p_{1}^{-1}\,$, that moves along a helical
polymer with a constant pitch $\frac{1}{2}v_{1}$.
We believe that the foregoing provides a precise and quantitative
dynamical underpinning for the conformon concept
put forward by Green and Ji and by others\cite{gree}.

We can go even further.
Let us first
consider the special case of a planar polymer,
by setting $\tau\equiv 0$ in the FW equations. We get
$\mathbf{t}_s=
\kappa\mathbf{N}\,,\,\mathbf{N}_s=-\kappa\mathbf{t}\,$, and
$\mathbf{B}=\mathbf{B}_c\,$, a constant vector.
Then, using the LL equation, a bit of algebra gives
$\mathbf{t}_u=\kappa_s\,\mathbf{B}_c$ and
$\mathbf{N}_u=(\kappa_{ss}/\kappa)\,\mathbf{B}_c$.
The non-extensibility condition
$\mathbf{t}_{su}=\mathbf{t}_{us}$ yields $\kappa_u=0\,$,
implying that $\kappa=\kappa(s)$.
The compatibility condition  $\mathbf{N}_{su}=\mathbf{N}_{us}$
gives, after the rescaling $s\rightarrow 2^{-1/2}s$, the equation
$\kappa_{ss}+\frac{1}{4}\kappa^{3}=E\,\kappa$ where $E$ is an
arbitrary constant. But this equation is
just the time-independent counterpart of
the time-dependent Schr\"odinger equation
$i\psi_u+\psi_{ss}-V\psi=0$
for a particle with wave function
$\psi=\kappa\,\exp\,(iEu)$, in the ``potential''
$V=-\frac{1}{4}\kappa^2(s)$.
Interestingly, it has been shown by Goldstone and
Jaffe\cite{goldstone} that the one-dimensional
Schr\"{o}dinger equation for a free particle in a thin tube
whose axis follows a space curve of curvature $\kappa$
acquires an effective attractive
potential $-\frac{1}{4}\kappa^2(s)$ (for small $\kappa$).
In our formalism, this has emerged naturally as the elastic energy of
the WLC. The time-independent Schr\"odinger equation above
has a (static) soliton solution
$\kappa=p\, {\rm sech}\,\,ps$ where $p=E^{1/2}$.
Combining these results, we have
$V=-\frac{1}{4}p^{2}\,{\rm sech}\,^{2}ps\,$, and
a corresponding localized wave function
$\psi= p\,{\rm sech}\,(ps)\,\exp\,(iEu)$. Applying this result to
an electronic wave function on a polymer, this suggests
the possibility that  a soliton, i.e., a nonlinear excitation
of the elastic polymer, can lead to electron localization.
In the non-planar case ($\tau\neq0$), the effective
potential continues to be  $V=-\frac{1}{4}\kappa^2$,
but now $\kappa = \kappa (s,u)$. In particular,
we can now interpret the wave
function $\psi$ as the one-soliton solution in the
time-dependent potential
$V=-\frac{1}{4}p_{1}^2\,{\rm sech}\,^{2}{\xi}_{1}$.
In the context of electron localization,
the electron gets trapped in a moving
potential well that travels along the polymer.
This ties in with the conformon concept as suggested
by Volkenstein\cite{volk} and by
Kemeny and Goklany\cite{keme}.
We conjecture
that this moving soliton plays a role in the
unexpectedly
high electrical conductivity found in biological membranes. \\

\noindent
{\it The WLRC model}:\,
Here a polymer such as DNA is represented as a slender
{\it twisted} elastic rod
with two  elastic constants that characterize
the bending energy and the twisting energy,
respectively\cite{fain}. The state of the polymer is
specified by the
configuration of the ribbon formed by a
longitudinal cross-section
of the rod containing its central axis.
The ribbon has a uniform twist angle
$\omega_{0}$ per unit
length around the central axis, to take into account
the double helical structure of DNA. In general,
the central axis of the ribbon is a space curve.
Let $\mathbf{t}$
be the unit tangent to
this axis at any point, and $\mathbf{N}_0$
the unit normal to the surface of the ribbon at that point.
Defining $\mathbf{B}_0=\mathbf{t}\times\mathbf{N}_0\,$,
the axis becomes a framed space curve on the
surface of the ribbon. It is obviously convenient to work with a
``co-twisted'' frame
$(\mathbf{t}, \mathbf{N}, \mathbf{B})$ where
$\mathbf{N}=\mathbf{N}_0\,\cos\,\omega_{0}s
+\mathbf{B}_0\,\sin\,\omega_{0}s$ and
$\mathbf{B}=
-\mathbf{N}_0\,\sin\,\omega_{0}s
+\mathbf{B}_0 \,\cos\,\omega_{0}s\,.$ In this frame
the axis of the polymer is a space curve {\it on the surface
of the ribbon}, and hence is described by the
Darboux-Ribaucour (DR) equations\cite{vais}
$ \mathbf{t}_s=\kappa_N\,\mathbf{N}+\kappa_g\,\mathbf{B}\,,\,
\mathbf{N}_s=-\kappa_N\,\mathbf{t}+\tau_g\,\mathbf{B}\,,\,
\mathbf{B}_s=-\kappa_g\,\mathbf{t}-\tau_g\,\mathbf{N}$.
Here $\kappa_N$ and $\kappa_g$ represent, respectively,
the normal and the geodesic curvatures, and $\tau_g$
is the geodesic torsion.
As the width of the ribbon goes to zero,
the geodesic curvature of any continuous
curve on its surface tends to zero. The only curve
which survives is the axis of
the ribbon, with curvature $\kappa_N$ and
torsion $\tau_g\,$. The DR equations reduce to
\beq
\mathbf{t}_s=\kappa_{N}\mathbf{N}\,,\,
\mathbf{N}_s = -\kappa_{N}\mathbf{t}+\tau_{g}\mathbf{B}\,,\,
{\rm and}\,\,
\mathbf{B}_s=-\tau_{g}\mathbf{N}\,,
\label{wlrc}
\eeq
where $\mathbf{N}$ and $\mathbf{B}$ now denote the normal and
binormal to the curve.
Once again, as described earlier in the case of the WLC model,
the frame $(\mathbf{t}, \mathbf{N}, \mathbf{B})$ can therefore be
regarded as rotating with a ``spatial angular velocity''
$\widetilde{\bm{\Omega}}
=\tau_{g}\mathbf{t}+\kappa_{N}\mathbf{B}$.
Hence the corresponding Hamiltonian is of the form
\beq
H= \frac{1}{2}\int(A\kappa_{N}^2+C\tau_{g}^2)\,ds
=\frac{1}{2}\int(A{\mathbf{t}_s}^2
+C{\mathbf{B}_s}^2)\,ds\,,
\label{wlrce}
\eeq
where $A$ and $C$ are constants. But this expression is precisely
the elastic energy of a twisted polymer in the
WLRC model\cite{fain,fn1}, where $A$ is the bending modulus as
before, and $C$ is the twist modulus.

We first discuss the symmetric case $A=C$. Using
Eq. (\ref{wlrc}) the (scaled) energy can be
written as   $H=\frac{1}{2}\int\mathbf{N}_s^2\,ds$.
We proceed as in the case of the WLC model. The constraint in this
case is $\mathbf{N}^2 = 1$. We finally obtain
\beq
\mathbf{N}_u=\mathbf{N}\times\mathbf{N}_{ss}
\label{LLN}
\eeq
as the counterpart of Eq. (\ref{LL}).
The discussion following the latter equation applies to the
present case, with $\mathbf{N}$ replacing $\mathbf{t}$.
Thus it is now $\mathbf{N}$ that behaves
like a spin vector field.

We drop the subscripts in $\kappa_N$ and $\tau_g$
hereafter, to simplify the notation.
Equations (\ref{wlrc}) and (\ref{LLN}) give
$\mathbf{N}_u=\tau_s\,\mathbf{t}+\kappa_s\,\mathbf{B}$.
This, along with the conditions
$\mathbf{N}\cdot\mathbf{t}=0\,,\,\mathbf{N}\cdot
\mathbf{B}=0$ and
$\mathbf{t}_{su}=\mathbf{t}_{us}\,$, yields
$\mathbf{t}_u=-\tau_s\,\mathbf{N}+
\frac{1}{2}(\kappa^2+\tau^2)\mathbf{B}$  and
$\mathbf{B}_u=
-\frac{1}{2}(\kappa^2+\tau^2)\mathbf{t}-\kappa_s\,\mathbf{N}.$
The compatibility condition $\mathbf{N}_{su}=\mathbf{N}_{us}$
then leads to\cite{radha}
\beq
\kappa_u+\tau_{ss}+\frac{\tau}{2}(\kappa^2+\tau^2)=0\,,\,\,
\tau_u-\kappa_{ss}-\frac{\kappa}{2}(\kappa^2+
\tau^2)=0\,.
\label{wlrcg}
\eeq
A short calculation shows that Eqs. (\ref{wlrcg}) represent
respectively, the real and imaginary parts of the NLS,
Eq. (\ref{nls}), if we set
$\psi=\kappa+i\tau$. In the WLC model, in contrast, it is
the function $\kappa\,\exp\,(i\int\tau\,ds)$
that satisfies the NLS.

For the one-soliton solution of the NLS
in Eq. (\ref{nlss}), the  energy density in WLRC model
once again has the form
$|\psi|^2=p_{1}^{2}\,{\rm sech}^{2}\,\xi_1\,$,
representing an
energy lump propagating with velocity $v_1$.
However, the curvature
and torsion differ from those in the WLC model.
They are now given by
 $\kappa=p_1\,{\rm sech}\,\xi_1\,\cos\,\xi_2$ and
$\tau=p_1\,{\rm sech}\,\xi_1\,\sin\,\xi_2$.
In contrast to the WLC model,
the curvature and torsion are
{\it both} localized functions,
with an additional spatial periodicity $\lambda=4\pi v_{1}^{-1}.$
This is the signature of a propagating localized
``supercoiled'' excitation (a conformon) in DNA.

Finally, we consider the general case $A\neq C$,
for which the scaled Hamiltonian is
$H=\frac{1}{2}\int(\kappa^2+\tilde{\tau}^2)\,ds$ where
$\tilde{\tau} =C\tau/A.$ The NLS is now satisfied by
the function
$\tilde{\psi}=\kappa+i\tilde{\tau}.$
The results for the one-soliton solution
described in the preceding paragraph continue to apply. The only
difference is that the torsion of the polymer is now scaled by a
factor $A/C$.

To conclude, we have provided an {\it exact}
analysis of the nonlinear
dynamics of elastic polymers in the
WLC and WLRC models using a
geometric formulation, and obtained several
new results. Although we have considered
only open polymers,
our analysis can be extended to
closed polymers as well. As a consequence of the
periodic boundary conditions in this case, the
solutions of the NLS
now turn out to be elliptic functions.

Based on our results we note the following
similarities between our soliton and the ``conformon''
conjectured in molecular biology\cite{gree, scot}: (i)
It is an energy
packet associated with a conformational strain
localized in a region
much shorter than the length of the molecule;
(ii) it needs a ``rigid
framework'' (i.e., a structure with finite bend/twist moduli)
for its formation;
and (iii) its  properties appear to depend on
the coupling between
{\it adjacent} bonds, e.g., in a
polypeptide chain. The compatibility of our results
with (i) and (ii) is obvious. As for (iii),
we note that in both the WLC and WLRC models,
the Hamiltonian density involves spatial derivatives
of unit vector fields of the form
${\mathbf w}_s^{2}$. But this is
just the continuum limit of the
nearest-neighbor interaction
${\mathbf w}_i\cdot{\mathbf w}_{i+1}$ where
$i$ is the site index on a discrete chain.
Further, as we have seen, ${\mathbf w(s)}$ acts like a spin
vector field. Thus the  dynamical evolution of the polymer is
governed by the nearest-neighbour (i.e., adjacent)  coupling of
an effective ``spin vector'' ${\mathbf w}_{i}$ at each site on the chain.
As is well known, there are several mechanisms
in condensed matter physics  which can lead to such effective spin
Hamiltonians. A secondary effect can also occur,
as we have discussed. As our soliton
(localized curvature) moves on the polymer, it can
trap an electron and drag
it along. This is in accordance with the
conformon concept of
Volkenstein\cite{volk}
and of Kemeny and Goklany\cite{keme}.
This  mechanism may be important in the understanding
of charge transport in biological molecules,
where curvatures abound. As we have demonstrated,
curved geometry and
nonlinearity are intimately related, and we believe
that our results are also of significance in
other kinds of transport phenomena in molecular biology.


\begin{thebibliography}{99}
\bibitem {fn1}Using Euler angles $(\theta,\phi,\psi)$ for
the rotation of the triad, we get
$\kappa^{2}_{N}=\theta^{2}_s+ \phi_s^2\,\sin^{2}\,\theta$
and $\tau_{g}^2=
(\phi_s\,\cos\,\theta +\psi_s)^2\,$,
where $\theta$ and $\phi$ determine
the position of $\mathbf{t}$ on the unit
sphere\cite{dando}.
\end{thebibliography}
\end{document}